\documentclass[sigconf]{acmart-me}

\AtBeginDocument{%
  \providecommand\BibTeX{{%
    \normalfont B\kern-0.5em{\scshape i\kern-0.25em b}\kern-0.8em\TeX}}}

\usepackage{algorithmic}
\usepackage{graphicx}
\usepackage{textcomp}
\usepackage{xcolor}
\usepackage{booktabs}
\usepackage{url}
\usepackage{multirow}
\usepackage{enumitem}
\usepackage{array}
\usepackage{hyperref}
\hypersetup{colorlinks,allcolors=black}

\usepackage[acronym, nowarn]{glossaries}
\makeglossaries

\newacronym{gi}{GI}{gastrointestinal}
\newacronym{ai}{AI}{artificial intelligence}
\newacronym{cadx}{CADx}{computer-aided diagnosis}
\newacronym{ml}{ML}{machine learning}
\newacronym{dl}{DL}{deep learning}
\newacronym{cnn}{CNN}{convolutional neural network}
\newacronym{crc}{CRC}{colorectal cancer}

\acmConference[MediaEval'20]{Multimedia Evaluation Workshop}{December 14-15 2020}{Online}
\setcopyright{rightsretained}
\copyrightyear{2020}
\acmYear{2020}
\acmDOI{}
\acmPrice{}
\acmISBN{}

\begin{document}
\title{Medico Multimedia Task at MediaEval 2020: \\Automatic Polyp Segmentation}

\author{Debesh Jha$^{1,2}$, Steven A. Hicks$^{1,3}$, Krister Emanuelsen$^{1}$, H{\aa}vard Johansen$^{2}$\\ Dag Johansen$^{2}$, Thomas de Lange$^{4,5,6}$, Michael A. Riegler$^{1}$, P{\aa}l Halvorsen$^{1,3}$}
\affiliation{%
 	\textsuperscript{1} SimulaMet, Norway \ \ \ \ \ \ \ \ \
    \textsuperscript{2} UiT The Arctic University of Norway  \ \ \ \ \ \ \ \ \ 
	\textsuperscript{3} Oslo Metropolitan University, Norway \\
	\textsuperscript{4} Augere Medical AS, Norway  \ \ \ \ \ \ \ \ \
	\textsuperscript{5} Sahlgrenska University Hospital, Sweden  \ \ \ \ \ \ \ \ \
	\textsuperscript{6} Bærum Hospital, Norway
}

\renewcommand{\shortauthors}{Jha et. al.}
\renewcommand{\shorttitle}{Medico Multimedia Task at MediaEval 2020}

\begin{abstract}
Colorectal cancer is the third most common cause of cancer worldwide. According to Global cancer statistics 2018, the incidence of colorectal cancer is increasing in both developing and developed countries. Early detection of colon anomalies such as polyps is important for cancer prevention, and automatic polyp segmentation can play a crucial role for this. Regardless of the recent advancement in early detection and treatment options, the estimated polyp miss rate is still around 20\%. Support via an automated computer-aided diagnosis system could be one of the potential solutions for the overlooked polyps. Such detection systems can help low-cost design solutions and save doctors time, which they could for example use to perform more patient examinations. In this paper, we introduce the 2020 Medico challenge, provide some information on related work and the dataset, describe the task and evaluation metrics, and discuss the necessity of organizing the Medico challenge.   
\end{abstract}

\maketitle 
\section{Introduction}\label{introduction}
The goal of \emph{Medico automatic polyp segmentation challenge} the benchmarking of polyp segmentation algorithms on new test images for automatic polyp segmentation that can detect and mask out polyps (including irregular, small or flat polyps) with high accuracy. The main goal of the challenge is to benchmark different computer vision and machine learning algorithms on the same dataset that could promote to build novel methods which could be potentially useful in clinical settings. Moreover, we emphasize on robustness and generalization of the methods to solve the limitations related to data availability and method comparison. The detailed challenge description can be found here \url{https://multimediaeval.github.io/editions/2020/tasks/medico/}. 

After three years of organizing the Medico Multimedia Task~\cite{riegler2017multimedia, pogorelov2018medico, hicks2019acm}, we present the fourth iteration in the series. With a focus on assessing human semen quality last year~\cite{hicks2019acm}, this year we build on the 2017~\cite{riegler2017multimedia} and 2018~\cite{pogorelov2018medico} challenges of automatically detecting anomalies in video and image data from the GI tract. We introduce a new task for automatic polyp \textit{segmentation}. In the prior \gls{gi} challenges, we classified the images into various classes. We are now interested in identifying each pixel of the lesions from the provided polyp images in this challenge. 
 
The task is important because \gls{crc} is the third most leading cause of cancer and fourth most prevailing strain in terms of cancer incidence globally~\cite{bray2018global}. Regular screening through colonoscopy is a prerequisite for early cancer detection and prevention of \gls{crc}. Regardless of the achievement of colonoscopy examinations, the estimated polyp miss rate is still around 20\%~\cite{kaminski2010quality}, and there are large inter-observer variabilities~\cite{mahmud2015computer}. An automated  \gls{cadx} system detecting and highlighting polyps could be of great help to improve the average endoscopist performance.

In recent years, \glspl{cnn} have advanced medical image segmentation algorithms. However, it is essential to understand the strengths and weaknesses of the different approaches via performance comparison on a common dataset. There are a large number of available studies on automatic polyp segmentation~\cite{fan2020pranet,guo2020polyp,jha2019resunet++,mahmud2020polypsegnet,jha2020doubleu,guo2019giana,wang2018development,jharealtime}. However, most of the conducted studies are performed on a restricted dataset which makes it difficult for benchmarking, algorithm development and reproducible results. Our challenge is utilizing the publicly available Kvasir-SEG dataset~\cite{jha2020kvasir}. The entire Kvasir-SEG dataset is used for training and an additional and unseen test dataset for benchmarking the algorithms.

In summary, the Medico 2020 challenge can support building future systems and foster open, comparable and reproducible results where \textit{the objective of the task is to find efficient solutions automatic polyp segmentation}, both in terms of pixel-wise accuracy and processing speed.

For the clinical translation of technologies, it is essential to design methods on multi-centered and multi-modal datasets. We have recently released several gastrointestinal endoscopy~\cite{pogorelov2017kvasir,pogorelov2017nerthus,borgli2020hyperkvasir}, wireless capsule endoscopy~\cite{smedsrud2020kvasir}, endoscopic instrument ~\cite{jhakvasirinstrument}, and polyp datasets~\cite{jha2020kvasir}. Thus, we have put in significant effort to address the challenges related to lack of public available datasets in the field of \gls{gi} endoscopy.

\begin{figure} [!t]
    \centering
    \includegraphics[height=1.6cm]{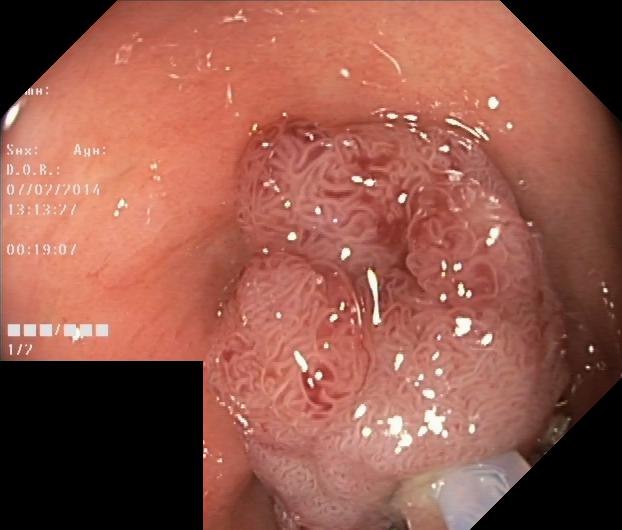}
    \includegraphics[height=1.6cm]{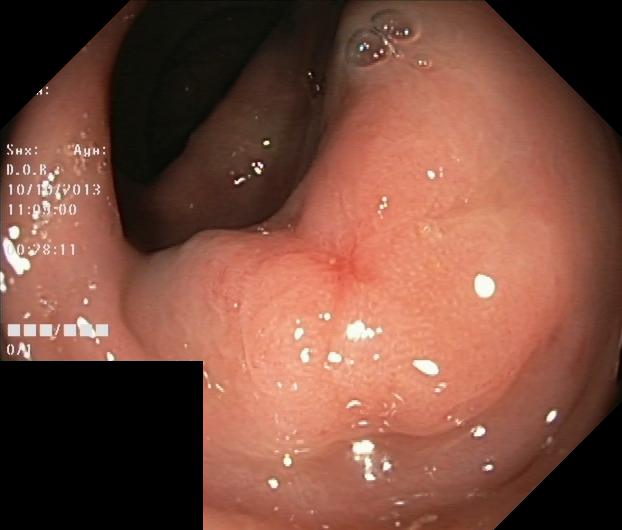}
    \includegraphics[height=1.6cm]{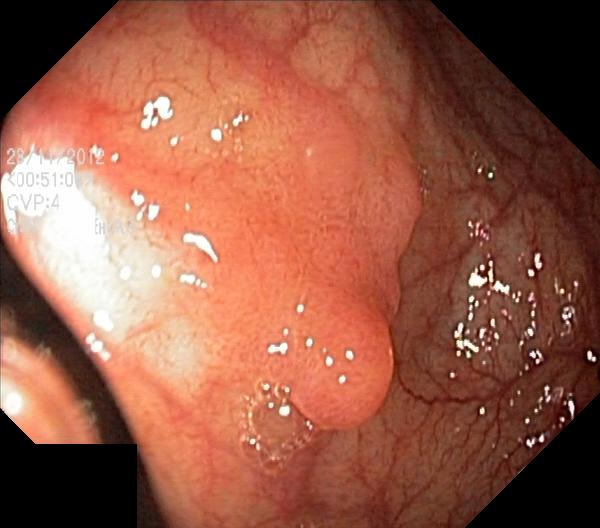}
    \includegraphics[height=1.6cm]{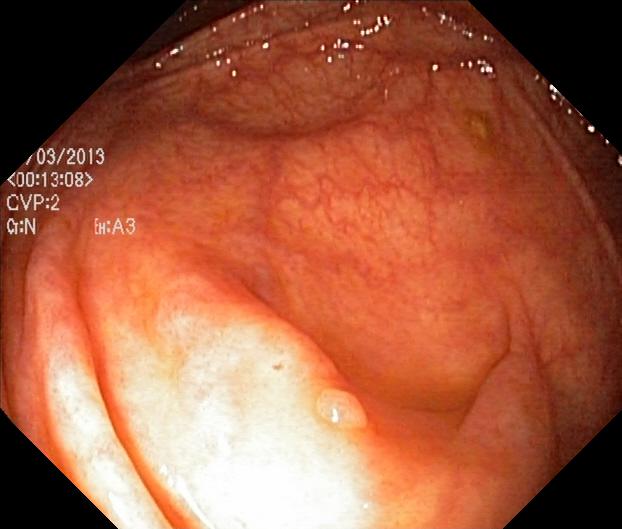}
    \includegraphics[height=1.6cm]{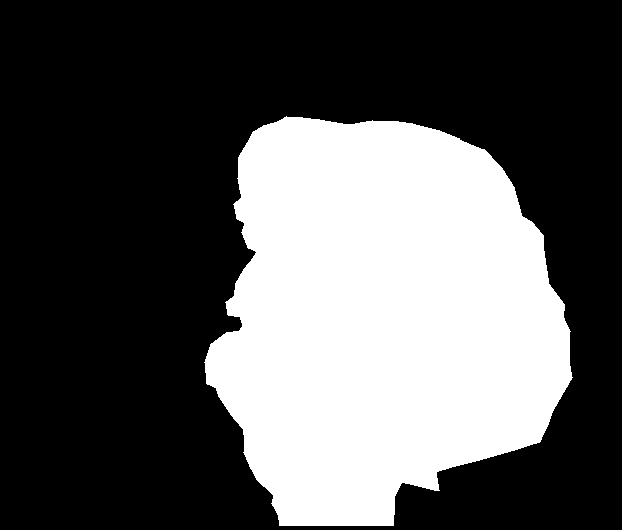}
    \includegraphics[height=1.6cm]{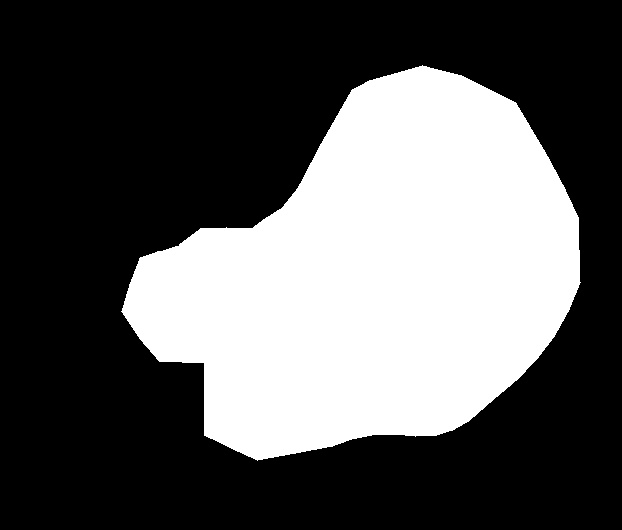}
    \includegraphics[height=1.6cm]{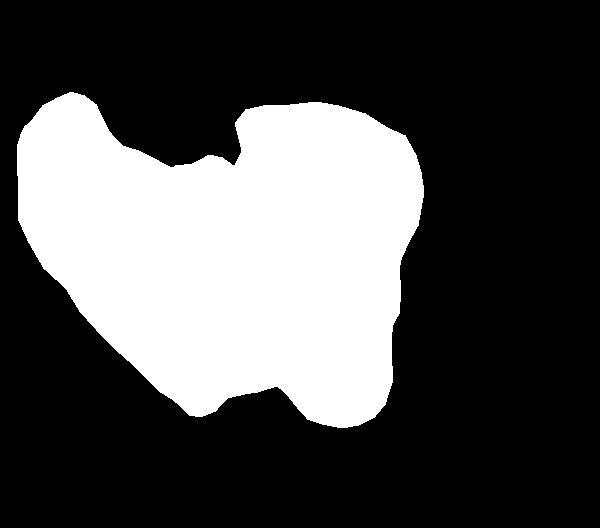}
    \includegraphics[height=1.6cm]{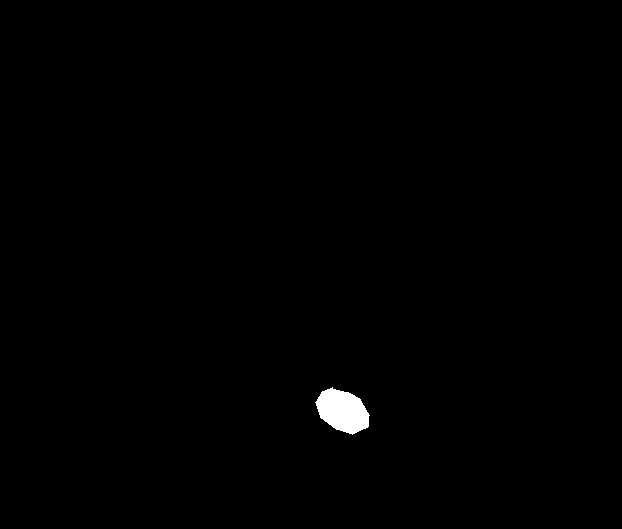}
    \caption{Polyps and corresponding masks from Kvasir-SEG}  
    \label{fig:images}
\end{figure}

\section{Dataset}\label{dataset}
\begin{figure} [!t]
    \centering
    \includegraphics[height=1.6cm]{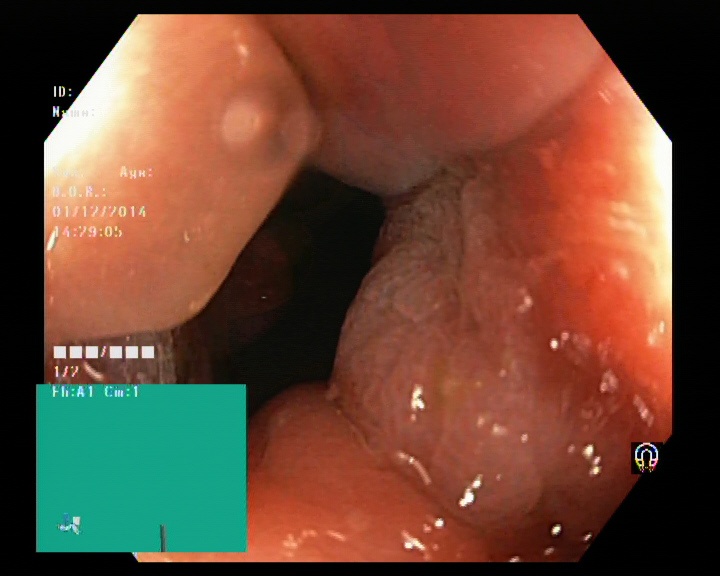}
    \includegraphics[height=1.6cm]{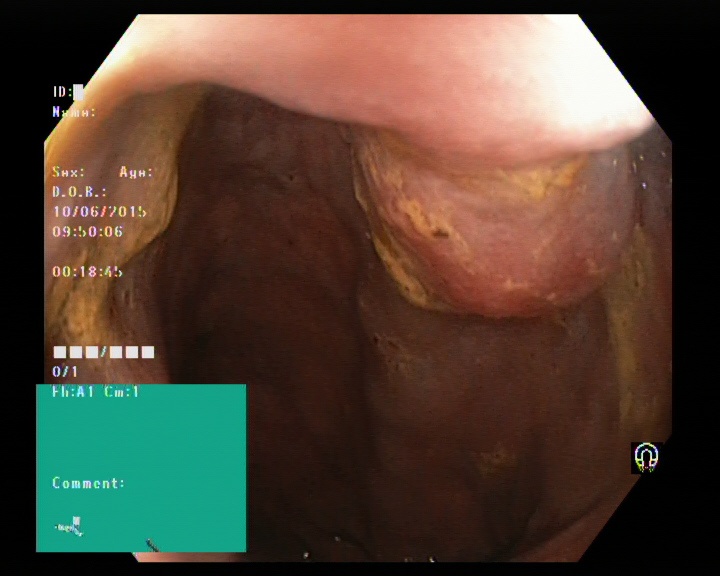}
    \includegraphics[height=1.6cm]{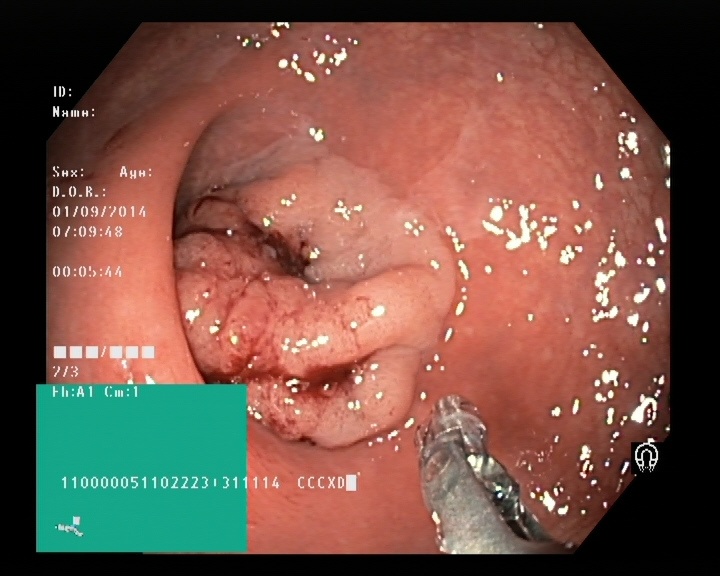}
    \includegraphics[height=1.6cm]{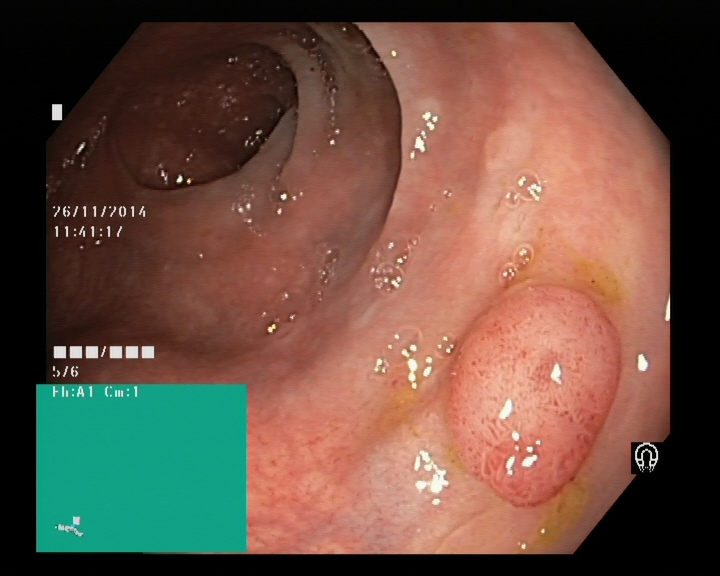}
    \includegraphics[height=1.6cm]{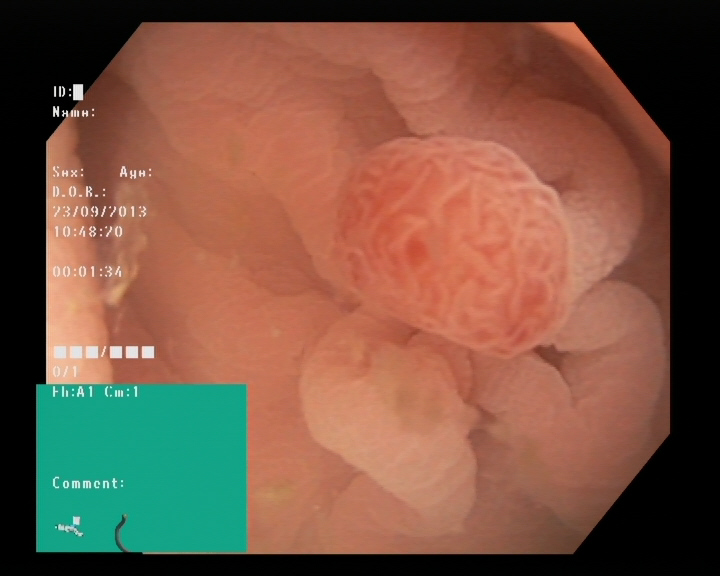}
    \includegraphics[height=1.6cm]{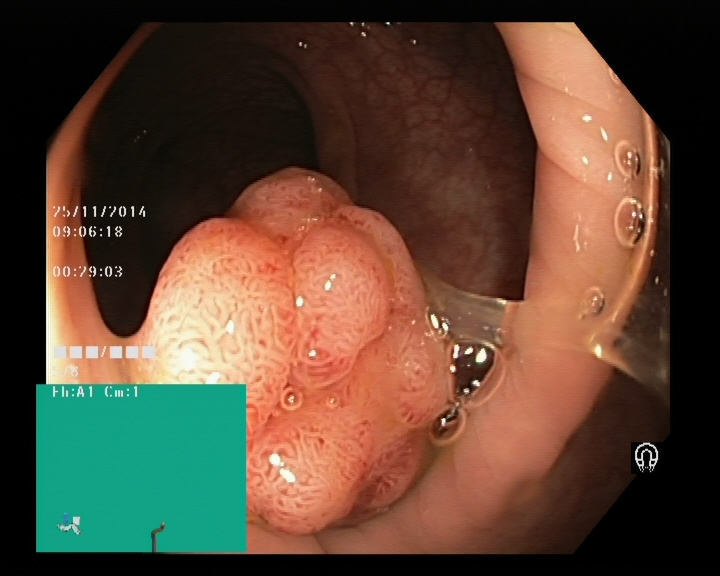}
    \includegraphics[height=1.6cm]{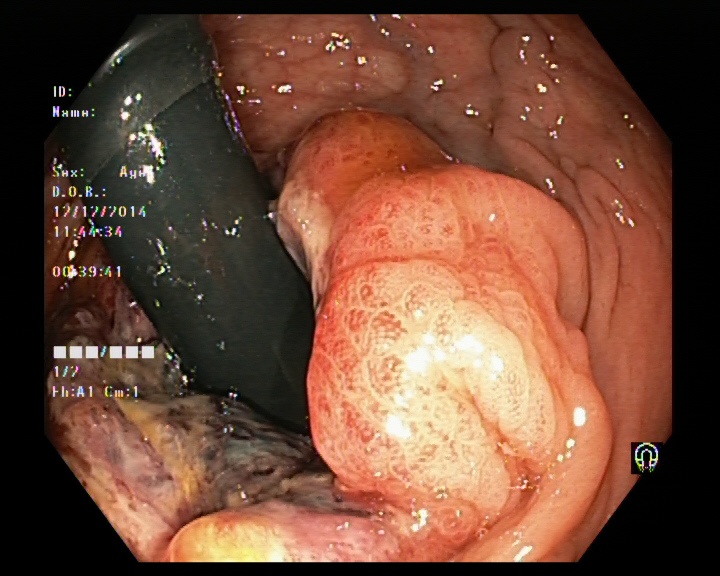}
    \includegraphics[height=1.6cm]{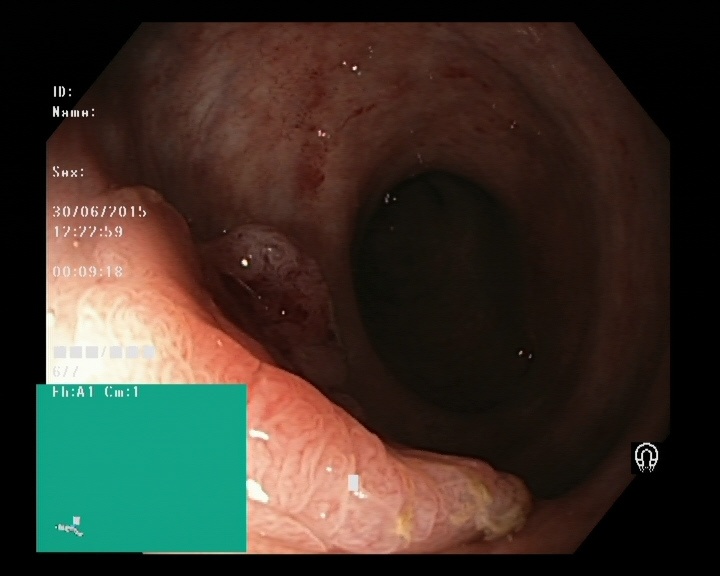}
    \caption{Examples polyps from the test images}  
    \label{fig:testimages}
\end{figure}

The Kvaris-SEG~\cite{jha2020kvasir} training dataset can be downloaded from \url{https://datasets.simula.no/kvasir-seg/}. It contains 1,000 polyp images and their corresponding ground truth mask as shown in Figure~\ref{fig:images}. The dataset was collected from real routine clinical examinations at B{\ae}rum Hospital in Norway by expert gastroenterologists. The resolution of images varies from $332 \times 487$ to $1920 \times 1072$ pixels. Some of the images contain a green thumbnail in the lower-left corner of the images showing the scope position marking from the ScopeGuide (Olympus) (see Figure~\ref{fig:testimages}). We annotate another separate dataset consisting of 160 new polyp images and use the resulting dataset as the test set to benchmark the participants' approaches. Figure~\ref{fig:testimages} shows some examples of test images used in the challenge.

\section{Task Description}
The participants are invited to submit their solutions for the two following tasks: segmentation and efficiency (speed).

\subsection{The automatic polyp segmentation task}
This task invites participants to develop new algorithms for segmentation of polyps. The main focus is to develop an efficient system in terms of diagnostic ability and processing speed and accurately segment the maximum polyp area in a frame from the provided colonoscopic images.  

There are several ways to evaluate the segmentation accuracy. The most commonly used metrics by the wider medical imaging community are the correct \textbf{Dice similarity coefficient (DSC)} or overlap index, and the \textbf{mean Intersection over Union (mIoU)}, also known as the Jaccard index. In clinical applications, the gastroenterologists are interested in pixel-wise detail information extraction from the potential lesions. The metrics such as DSC and mIoU are used to compare the pixel-wise similarity between the predicted segmentation maps and the original ground truth of the lesions. 

The DSC is a metric for comparison of the similarities between two given samples. If \textit{tp}, \textit{tn}, \textit{fp}, and \textit{fn} represent the number of true positive, true negative, false positive and false negative  per-pixel predictions for an image, respectively,  then the DSC is given as
\begin{equation*}\label{eq:dice}
\text{DSC} = \frac{2 \cdot tp} {2 \cdot tp + fp + fn}
\end{equation*}
Furthermore, the IoU  is then defined as the ratio of intersection of two metrics over a union of two corresponding metrics.  The mean IoU computes IoU of each semantic class of an image and calculate the mean over each classes. The IoU is defined as:
\begin{equation*}\label{eq:iou}
\text{IoU} = \frac{tp} {tp + fp + fn}
\end{equation*}

Moreover, in the polyp image segmentation task (i.e., a binary segmentation task), \textbf{precision} (positive predictive value) shows over-segmentation, and \textbf{recall} (true positive rate) shows under-segmentation. Over-segmentation means that the predicted image covers more area than the ground truth in some part of the frame. The under-segmentation implies that the algorithm has predicted less polyp content in some portion of the image compared to its corresponding ground truth. We also encourage participants to calculate precision and recall, and these are given by: 

\begin{equation*}
\text{Precision} =\frac{tp} {tp + fp}
\end{equation*}
\begin{equation*}
\text{Recall} = \frac{tp} {tp + fn}.
\end{equation*}

The main metric for evaluation and ranking of the teams is \textbf{mIoU}. There is a direct correlation between mIoU and DSC. Therefore, we have only used one metric. If the teams have the same mIoU values, then the teams will be further evaluated on the basis of the higher value of the DSC. For the evaluation, we ask the participants to submit the predicted masks in a zip file. The resolution of the predicted masks must be equal to the test images. 

\subsection{The algorithm speed efficiency task}
Real-time polyp detection is required for live patient examinations in the clinic. It can gain gastroenterologist attention to the region of interest. Thus, we also ask participants to participate in the efficiency task. The algorithm efficiency task is similar to the previous task, but it puts a stronger emphasis on the algorithm's speed in terms of frames-per-second. 

Submissions for this task will be evaluated based on both the algorithm's speed and segmentation performance. The segmentation performance (the segmentation accuracy) will be measured using the same \textbf{mIoU} metric as described above for the first task, whereas speed will be measured by \textbf{frames-per-second (FPS)} according to the following formula:
\begin{equation*}\label{fps}
FPS = {\frac{\#frames} {sec}}
\end{equation*}
For this task, we require participants to submit their proposed algorithm as part of a Docker image so that we can evaluate it on our hardware. We evaluate the performance of the algorithm on the Nvidia GeForce GTX 1080 system. For the team ranking, we set a certain  mIoU as threshold for considering it as a valid efficient segmentation solution and rank according to the FPS. 
\vspace{-2mm}
\section{Discussion and Outlook}\label{Discussion}
Currently, there is a growing interest in the development of \gls{cadx} systems that could act as a second observer and digital assistant for the endoscopists. Algorithmic benchmarking is an efficient approach to analyze the results of different methods. A comparison of different approaches can help us to identify challenging cases in the data. We then can discriminate the image frames into simple, moderate, and challenging images. Later on, we can target to develop models on the challenging images that are usually missed out during a routine examination to design better \gls{cadx} systems. We hope that this approach would help us to design better performing algorithms/models that may increase the efficiency of the health system. 

\bibliographystyle{ACM-Reference-Format}
\bibliography{references}
\end{document}